\documentclass[journal = jacsat,manuscript=letter,layout=twocolumn]{achemso}

\usepackage{graphicx} % Required for inserting images
\usepackage[utf8]{inputenc}
\usepackage{array}
\usepackage{booktabs}
\usepackage{hypdoc}
\usepackage{listings}
\usepackage{lmodern}
\usepackage{mathpazo}
\usepackage{microtype}

\usepackage{caption}
\usepackage{float}
\usepackage{geometry}
\usepackage{natbib}
\usepackage{setspace}
\usepackage{xkeyval}

\usepackage{multirow}

\def \PW{Department of Semiconductor Materials Engineering, Faculty of Fundamental Problems of Technology, Wrocław University of Science and Technology, Wybrzeże Wyspiańskiego 27, 50-370 Wrocław, Poland}
\def \IFT{Institute of Theoretical Physics, Faculty of Physics, University of Warsaw, Pasteura St. 5, 02-093 Warsaw, Poland}

\author{Kamil Wrzos}\affiliation{\PW}
\author{Magdalena Birowska}\affiliation{\IFT}
\author{Miłosz Rybak}\affiliation{\PW}
\email{milosz.rybak@pwr.edu.pl}

\title{Symmetry-Breaking Phenomena in MnPS$_3$/TMDC Heterostructures: Non-relativistic Spin Splitting, Altermagnetism and Spin-Valley Effects} 

\begin{document}

\date{\today} 

\begin{abstract}

We explore symmetry-breaking phenomena in MnPS$_3$/TMDC 
(MoS$_2$, WS$_2$, MoSe$_2$, WSe$_2$) heterostructures using first-principles calculations, considering two high-symmetry stacking configurations, S1 and S2, which differ not only by their interfacial registry but also by a 30$^\circ$ twist between the layers. Depending on the stacking geometry, the systems exhibit two distinct types of nonrelativistic spin splitting (NRSS): S2 hosts altermagnetic-like band crossings, while S1 shows global spin splitting characteristic of symmetry-breaking NRSS. Magnetic exchange and anisotropy parameters indicate that the intrinsic magnetic properties of MnPS$_3$ are largely preserved upon interfacing.  Including spin–orbit coupling, we find tunable conduction-valley splitting controlled by the MnPS$_3$ spin orientation.  Our results identify MnPS$_3$ as a symmetry-tunable antiferromagnetic substrate capable of inducing and controlling spin and valley effects in 2D heterostructures without relying on net magnetization or strong SOC, offering a route toward nonvolatile valleytronic functionalities.

\end{abstract}

\section{Introduction}

\begin{figure}
\label{Fig0}
\includegraphics[width=0.5\textwidth]{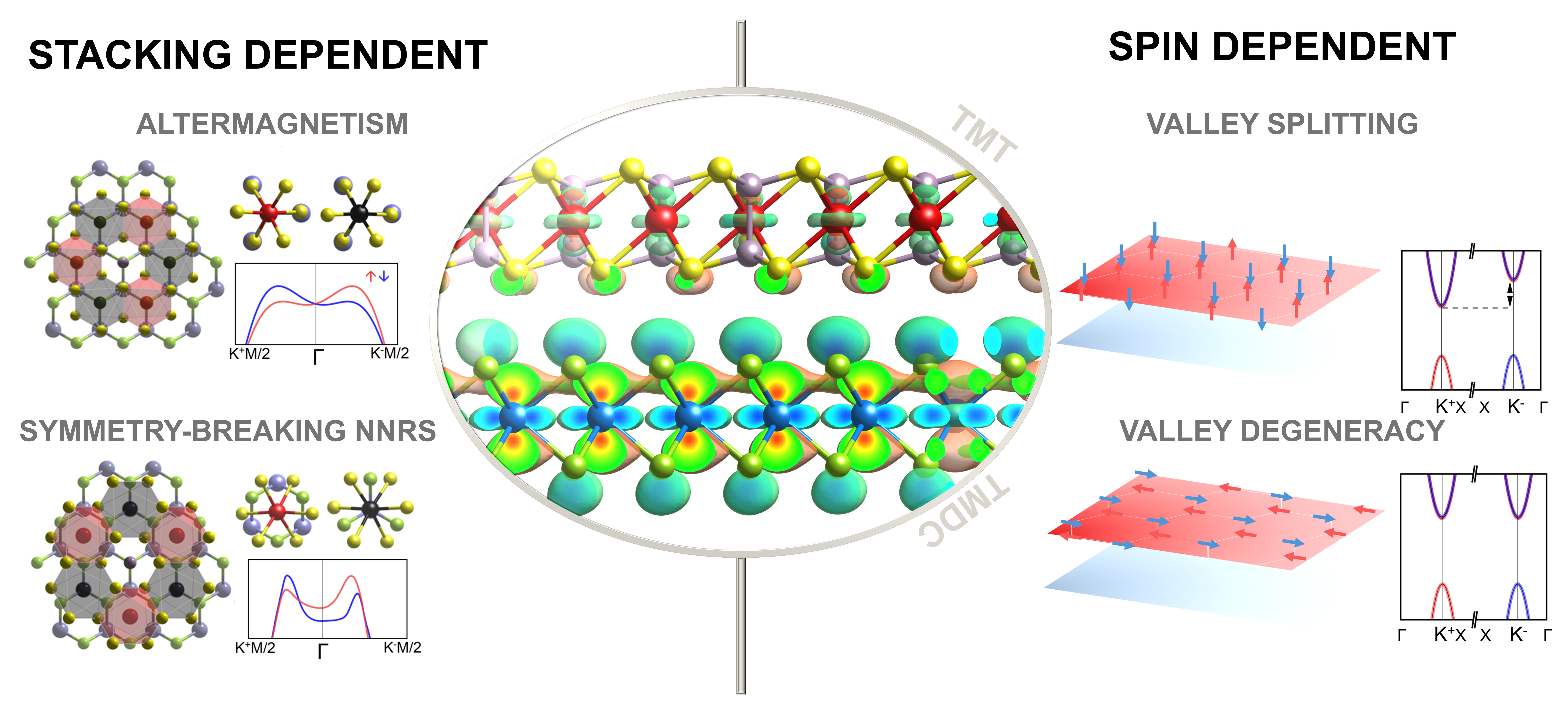}
\end{figure}

\begin{figure*}
\centering
\includegraphics[width=1.02\textwidth]{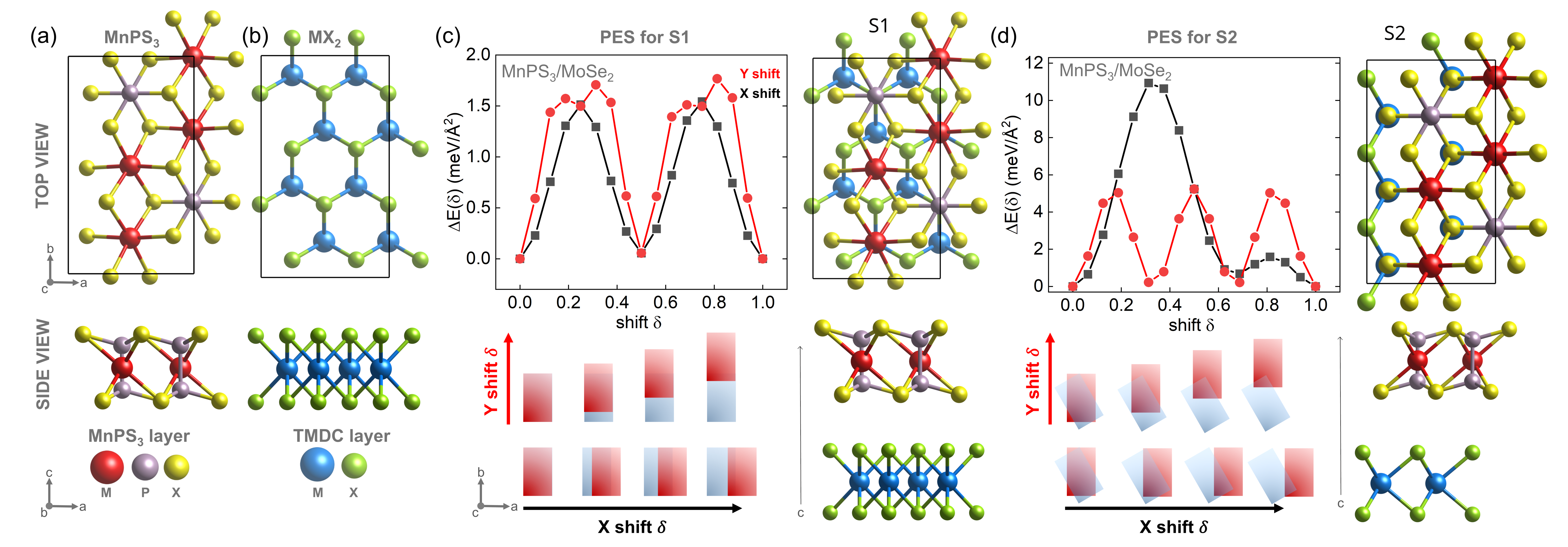}
\caption{Interlayer potential energy surface (PES) for the representative system of MnPS$_3$/MoS$_2$. Top and side views of (a) pristine TMT (MnPS\textsubscript{3}) layer, (b) TMDC MX\textsubscript{2} (M = Mo, W and X = S, Se) layer.
In each case, rigid in-plane shift of the MnPS$_3$  relative to the  TMDC layer are performed along the crystallographic x and y directions, quantified by a relative displacement parameter $\delta$. The PES is defined as $\Delta E(\delta)= (E(\delta)-E(\delta=0))/A_{hBL}$, where $A_{hBL}$ is the surface area of heterobilayer. All scans are rigid lateral translations at fixed interlayer spacing, where no in-plane or out-of-plane atomic relaxations are allowed-so the PES is registry-dependent interlayer potential. Schematic representations of the translation paths are presented at the bottom of (c) and (d) panels, whereas the right side of these  panels show the reference stackings S1 ($\delta=0$) and S2 ($\delta=0$), for which the minima of PES occur. }
\label{Fig1}
\end{figure*}

Two-dimensional (2D) van der Waals (vdW) heterostructures have emerged as a versatile platform for exploring novel physical phenomena arising from interlayer coupling, symmetry breaking, and tunable electronic correlations \cite{PhysRevB.108.075416, Zhang2025,Sierra2021,Masseroni2024}. By vertically stacking distinct 2D crystals, one can engineer artificial materials with tailored functionalities that are not present in the individual constituents. Transition metal dichalcogenides (TMDCs), in particular, have attracted considerable attention due to their valley-selective optical transitions, and semiconducting character, making them ideal components for opto-spintronic \cite{doi:10.1021/acs.nanolett.7b01393,Ahn2020} and valleytronic devices \cite{Tyulnev2024,Zhang2021,Robert2021}.  

In parallel, 2D magnetic materials such as MnPS$_3$ have gained prominence as model systems for exploring magnetism in the monolayer limit \cite{Strasdas2023,Autieri2022,PhysRevB.109.054426}. MnPS$_3$ is an antiferromagnetic (AFM) semiconductor belonging to the MPX$_3$ family, characterized by a honeycomb lattice of Mn$^{2+}$ ions coupled via superexchange through sulfur and phosphorus ligands. Its weak interlayer coupling, robust magnetic order, and easily accessible exfoliation make it an ideal substrate for spin- and symmetry-dependent proximity effects in vdW heterostructures \cite{PhysRevResearch.4.023256}. Recent theoretical and experimental works have demonstrated that interfacing TMDCs with magnetic layers can induce exchange fields, modify valley polarization, and even break time-reversal symmetry without introducing significant structural distortions or external magnetic fields \cite{Beer2024,Seyler2018}.  

Despite these advances, the microscopic mechanisms governing spin and valley interactions in AFM-based heterostructures are still not fully understood. In particular, the role of stacking geometry, interlayer hybridization, and strain in controlling symmetry breaking and band alignment has not been systematically explored. Moreover, while nonrelativistic spin splitting (NRSS) and altermagnetism have recently been identified as key manifestations of spin symmetry breaking in collinear antiferromagnets \cite{Yuan_NRSS,Fedchenko2024,PhysRevLett.132.036702}, their realization and tunability in realistic 2D heterostructures remain open questions. Understanding how stacking order and magnetic orientation modulate these effects is essential for designing controllable spin-valley systems that operate without net magnetization or strong spin--orbit coupling.  

In this work, we perform a comprehensive first-principles study of MnPS$_3$/TMDC (MoS$_2$, WS$_2$, MoSe$_2$, WSe$_2$) heterostructures to uncover the interplay between stacking geometry, strain, magnetism and optical properties. We focus on two high-symmetry stacking configurations, S1 and S2, which represent distinct atomic registries at the interface. Our findings identify MnPS$_3$ as a symmetry-tunable substrate for TMDC layers. By engineering stacking geometry at the interface and controlling the spin orientation in MnPS$_3$, it is possible to access multiple symmetry regimes-including altermagnetic-like states-and achieve spin-controlled valley selection without relying on net magnetization or strong spin–orbit coupling. These results position MnPS$_3$/TMDC heterostructures as promising platforms for next-generation nonvolatile valleytronic and opto-spintronic devices.

\section{Computational Details}
We performed DFT spin-polarized calculation with PBE generalized gradient exchange-correlation functional \cite{Perdew1996} with DFT-D3 dispersion correction scheme \cite{Grimme2010}.
To account for the half-filled electronic configuration \cite{Autieri2022} a Hubbard onsite Coulomb parameter of U = 1.8 eV was applied within the Dudarev formalism \cite{Dudarev1998} in agreement with $\mu$-ARPES observations  \cite{Strasdas2023}. A vacuum spacing of 25 \AA~was introduced along the out-of-plane direction to suppress spurious interactions between periodic images of the heterostructure.
Full geometry optimization was performed with a force convergence criterion of  $10^{-2}$ eV \AA$^{-1}$ per atom and a plane-wave energy cutoff of 500 eV. 
For atomic relaxation and self-consistent calculations regular Monkhorst–Pack grids of 5×9×1 were used \cite{Monkhorst1976}.

By performing scf computation without SOC it is possible to implement parametrization of nonrelativistic spin splitting (NRSS) by calculating its average splitting ($\langle \mathrm{AS} \rangle$) using the formula:

\begin{equation}
    \langle AS \rangle = \sqrt{\frac{1}{N_{occ}N_\mathbf{k}}\sum_{\mathbf{k},n}(\epsilon_{n,\mathbf{k},\uparrow}-\epsilon_{n,\mathbf{k},\downarrow})^2},
\end{equation}

where $N_{occ}$ is the number of occupied bands by 3d Mn electrons (in our case 5 electrons per atom, so 20 in total), $N_\mathbf{k}$ number of k-points, and $\epsilon_{n,\mathbf{k},\uparrow/\downarrow}$ is energy of $n$-band, in $\mathbf{k}$-point, of $\uparrow/\downarrow$ spin.
For normalization purposes ($N_{occ}$) we only considered 3d Mn electrons as they are the main contribution to NRSS due to their half-filling resulting in creation of a strongly correlated electronic system.

\section{Results}

We begin our calculations by constructing two distinct high-symmetry stacking configurations of MnPS$_3$/TMDCs heterostructures, denoted as the stacking S1 and  stacking S2, as illustrated in Fig. \ref{Fig1}(c) and (d), respectively. 

The lattice constants of the TMDCs in 2H  structural phase are nearly half of MnPS$_3$ primitive cell (see Table \ref{tab1}). To perform the band structure calculations of the resulting heterobilayer, we constructed the smallest rectangular supercell (see Fig. \ref{Fig1}a) that accommodates these two stackings and various spin magnetic phases. Because the MnPS$_3$ and TMDC lattices are incommensurate within the chosen supercell of heterobilayer, a lattice mismatch of each layer is summarized in Table \ref{tab1}.  Each stacking configuration was fully relaxed, including both ionic positions and cell parameters ($a_{hBL}$ and $b_{hBL}$).

For both stacking configurations, each constituting layer is strained. 
The MnPS$_3$ sheet exhibits the larger strain, with $|\varepsilon_{MnPS_3}| \approx 3.0$–$5.9\%$, compared with $|\varepsilon_{TMDC}|\approx1.0$–$4.2\%$ for the TMDC layer (Table \ref{tab1}). This follows from the lower in-plane stiffness of MnPS$_3$, particularly its 2D Young’s modulus as reported in \cite{} $\sim 60$–$120~\mathrm{N/m}$ \cite{https://doi.org/10.1002/adma.201707433}, whereas for MoS$_2$, WS$_2$, WSe$_2$ it is $\sim 160$-$180~\mathrm{N/m}$ \cite{Liu2014, CUI20222975}. In other words, MnPS$_3$ is a softer material and therefore accommodates larger strain than TMDCs. To minimize elastic energy (i.e.,  overall lattice mismatch $\varepsilon_{hBL}$, see Table \ref{tab1}), S1 is preferable for sulfide heterostructures $\varepsilon_{hBL}\approx$4\% in S1 vs $\varepsilon_{hBL}\approx$10\% in S2), whereas S2 is preferable for selenides. Because Se-based TMDCs have larger lattice constants, S2 exhibits modest TMDC tension ($\varepsilon_{TMDC}\approx$ 1–2.5\%) and moderate MnPS$_3$ compression ($\varepsilon_{MnPS_3}\approx$3–4\%), while S1 would require $\varepsilon_{MnPS_3}\approx$6\% tensile strain in MnPS$_3$. For sulfides, the roles reverse. Namely, S2 would impose $\varepsilon_{TMDC}\approx$4\% tension on the much stiffer TMDC, incurring a higher elastic cost.

\begin{table*}[t]
    \caption{Structural parameters and band offsets of various MnPS$_3$/TMDC heterobilayers. The $a_{hBL}$ and $b_{hBL}$ are optimized lattice constants of heterobilayers. The optimized parameters of pristine monolayers ($a_{ML}$) are MnPS$_3$ (6.064 \AA{}), MoS$_2$ (3.167\AA{}), WS$_2$(3.176\AA{}), MoSe$_2$ (3.299\AA{}), WSe$_2$ (3.294\AA{}), where the $b_{ML}=\sqrt{3}*a_{ML}$. The lattice mismatch ($\varepsilon$) of each layer is quantified by $\epsilon_{ML}=\frac{a_{hBL}-a_{ML}}{a_{ML}} \times 100\%$, and $\varepsilon_{hBL}$ is the overall lattice mismatch expressed as $\varepsilon_{hBL}=|\varepsilon_{TMDC}|+|\varepsilon_{MnPS_3}|$, where positive and negative signs indicate the tensile and compressive strain, respectively. The band offsets are defined as VBO$=$VBM$_{TMDC}-$VBM$_{MnPS_3}$, CBO$=$CBM$_{TMDC}-$CBM$_{MnPS_3}$, in eV. The \textit{band align.} label denotes the band-alignment type of the heterostructure. Namely, type-II occurs when both the VBO and CBO are positive,  while type-I arises when VBO is positive and CBO is negative.}
    \centering
    \resizebox{1.0\textwidth}{!}{%
    \begin{tabular}{cccccccccc}
        \toprule
        System & Stacking & 
        $a_{hBL}$ (\AA) & $b_{hBL}$ (\AA) & 
        $\varepsilon_\mathrm{TMDC}$ (\%) & 
        $\varepsilon_\mathrm{MnPS_3}$ (\%) & $\varepsilon_{hBL}$ (\%) &
        VBO (eV) & CBO (eV) & band align.\\
        \midrule
        \multirow{2}{*}{MnPS$_3$/MoS$_2$} & S1 & 6.245 & 10.817 & -1.4 &  3.0 & 4.4 & 0.16 & 0.29 & type II \\
                                          & S2 & 5.714 &  9.895 &  4.2 & -5.8 & 10.0&0.02 & -0.73 & type I\\
                                          \hline
        \multirow{2}{*}{MnPS$_3$/WS$_2$}  & S1 & 6.262 & 10.847 & -1.4 &  3.3 & 4.7&0.34 & 0.42 & type II\\
                                          & S2 & 5.715 &  9.898 &  3.9 & -5.8 &9.7  &0.20 & -0.49 &type I\\
                                            \hline
        \multirow{2}{*}{MnPS$_3$/MoSe$_2$}& S1 & 6.417 & 11.114 & -2.7 &  5.8 & 8.5 &0.63 & 0.46 & type II\\
                                          & S2 & 5.857 & 10.402 &  1.0 & -3.4 & 4.4 &0.16 & -0.05 & type I\\
                                            \hline
        \multirow{2}{*}{MnPS$_3$/WSe$_2$} & S1 & 6.421 & 11.122 & -2.5 &  5.9 & 8.4 &1.01 & 0.54 & type II\\
                                          & S2 & 5.847 & 10.127 &  2.5 & -3.6 & 6.1 &0.37 & 0.17 & type II\\

        \bottomrule
    \end{tabular}%
    }
    \label{tab1}
\end{table*}

\paragraph{Registry-dependent interlayer potential.}
We also examine other stacking configurations generated from the reference structures S1 and S2 by introducing lateral displacements  $\delta$ along the x  and y  crystallographic directions. 
The energy landscapes presented in Fig. \ref{Fig1}(c, d) illustrate the cost of laterally shifting MnPS$_3$ over MoSe$_2$. High-energy registries (maxima in the graphs) correspond to stackings where atoms experience stronger Pauli repulsion,  due to unfavorable on top chalcogen–chalcogen wavefunction overlap at the interface. Minima occur for reference stackings S1 ($\delta=0$) and S2 ($\delta=0$), where the interfacial chalcogen atoms are positioned closer to hollow-like sites (see side views of hBL - bottom right in Figs. \ref{Fig1}(c, d)). Van der Waals forces act as the background adhesion between the layers and are most effective in these hollow configurations, making them energetically preferred. 
In particular, for  S1 reference stacking, the PES is shallow, with the energy barrier of $\sim 2$ meV/\AA$^2$.
The red ($Y$) and black ($X$) curves exhibit a repeating pattern consistent with the lattice, with an approximately two-fold periodicity.
The barriers are similar-about $1.5$ meV/\AA$^2$ along $X$ and $1.8$ meV/\AA$^2$ along $Y$-indicating a largely isotropic landscape with only minor directional differences.
Note that equal fractional shifts $\delta_x$ and $\delta_y$ do not correspond to equal real-space distances because the lattice vectors
have different lengths ($L_x = a$, $L_y = \sqrt{3}\,a$). Consequently, analogous atomic registries are reached at different
values of $\delta_x$ and $\delta_y$.

In the S2 stacking, the interlayer potential is strongly anisotropic. Particularly, along $X$ (black) PES  is deep, with a dominant sliding barrier of $\approx$ 12 meV/\AA$^2$ (peak near $\delta \approx 0.4$), whereas along $Y$ (red) it is much weaker $7$ meV/\AA$^2$ and follows a different periodic motif. This contrast arises because $X$ and $Y$ pass through distinct interfacial registries. Namely, the $X$ path encounters on-top chalcogen–chalcogen overlapping that maximize short-range Pauli repulsion, while the $Y$ path largely avoids such configurations. Consequently, S2 exhibits registry locking-a deep, highly anisotropic potential that pins the bilayer against $X$-directed sliding while allowing comparatively softer motion along $Y$.

Overall, our results reveal that stacking S1 produces a smoother and only weakly corrugated interlayer potential, which facilitates stronger interlayer decoupling, lowers the sliding barrier, and reduces friction, thereby enabling easier relative motion. In contrast, configuration S2 gives rise to a deeper and strongly anisotropic potential-energy landscape that favors registry locking and results in higher static friction-particularly stiff along the X direction but comparatively softer along Y.

\begin{figure}
\centering
\includegraphics[width=\linewidth]{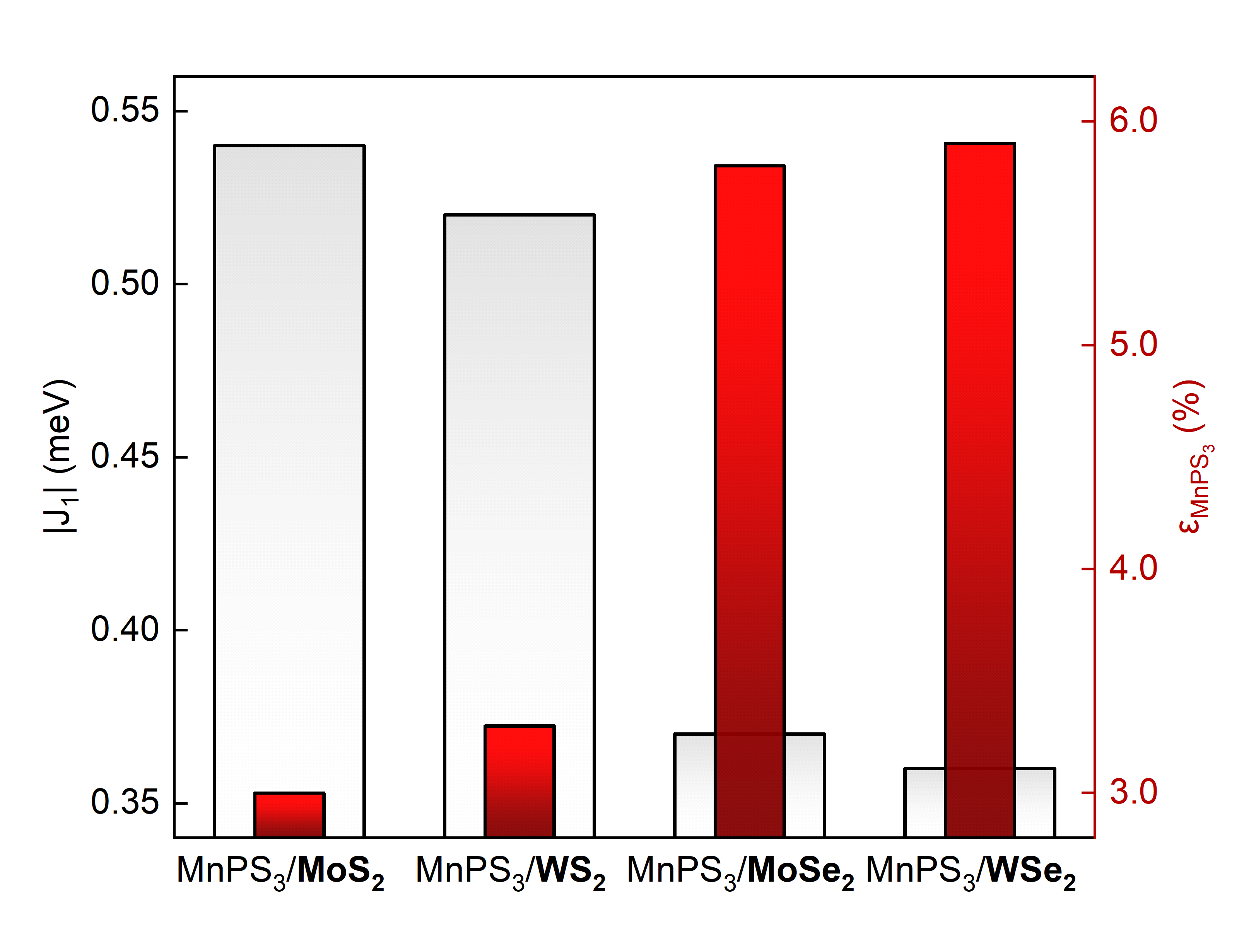}
\caption{Heisenberg exchange parameters ($|J_1|$) in meV and MnPS$_3$ strain ($\varepsilon_{MnPS_3}$) in percent for MnPS$_3$/TMDC heterostructures (S1) in various stacking configurations.}
\label{Fig5}
\end{figure}

\paragraph{Strain-driven modulation of exchange couplings in MnPS$_3$.}

To quantify the magnetic interactions in MnPS$_3$/TMDC heterostructures, we employed the extended spin Hamiltonian:

\begin{equation}
H = -\frac{1}{2} \sum_{ij} J_{ij}\mathbf{S}i \cdot \mathbf{S}j
- \frac{1}{2} \sum_{ij} \lambda_{ij}S_i^z S_j^z
- A \sum_i (S_i^z)^2 ,
\end{equation}

where $J_{ij}$ denotes the isotropic Heisenberg exchange, $\lambda_{ij}$ anisotropic exchange constants, and $A$ the on-site anisotropy (magnetocrystalline anisotropy energy, MAE), with $z$ taken as the global spin quantization axis. In our model we considered exchange interaction up to the next-next-nearest neighbor, therefore we introduced exchange parameters $J_i, \lambda_i$, where $i=1,2,3$. The parameters were obtained by fitting total-energy differences from non-collinear DFT calculations including SOC.

The extracted values ($J_i$, $\lambda_i$, $A$) are summarized in Supplementary Information in Table S2, where the parameters obtained for the pristine MnPS$_3$ are in line with previous reports \cite{PhysRevResearch.4.023256,Autieri2022}.  Comparison with pristine MnPS$_3$  shows that the heterostructures introduce moderate modifications of the exchange couplings $J_i$. The magnitude of the dominant $J_1$ coupling decreases as the overall lattice mismatch ($\varepsilon_{hBL}$) and tensile strain in MnPS$_3$
 increase, as illustrated in Fig.~\ref{Fig5}. Tensile strain in MnPS$_3$ weakens the nearest-neighbor AFM exchange due to the larger separation between magnetic ions. Conversely, compressive strain is expected to enhance the exchange interaction, thereby increasing the Néel temperature. Control calculations for a strained MnPS$_3$ monolayer, corresponding to the strain present in the MnPS$_3$/MoS$_2$ heterobilayer, yield nearly identical exchange parameters to those of the heterostructure. This indicates that the changes originate primarily from strain rather than from proximity effects (see Fig.~\ref{Fig5}). The other parameters ($\lambda_i$, $A$) are on the order of $\mu$eV and thus negligible, which is consistent with MnPS$_3$ being often described as a two-dimensional Heisenberg AFM with very weak easy-axis anisotropy.
Overall, these results demonstrate that the intrinsic Heisenberg-like antiferromagnetism of MnPS$_3$ is largely preserved in the MnPS$_3$ /TMDC heterostructures, with only weak strain-induced modifications and negligible dependence on stacking. For consistency, our analysis was restricted to three S1 translations and did not include  S2 stacking.
\begin{figure*}\centering
\includegraphics[width=1\textwidth]{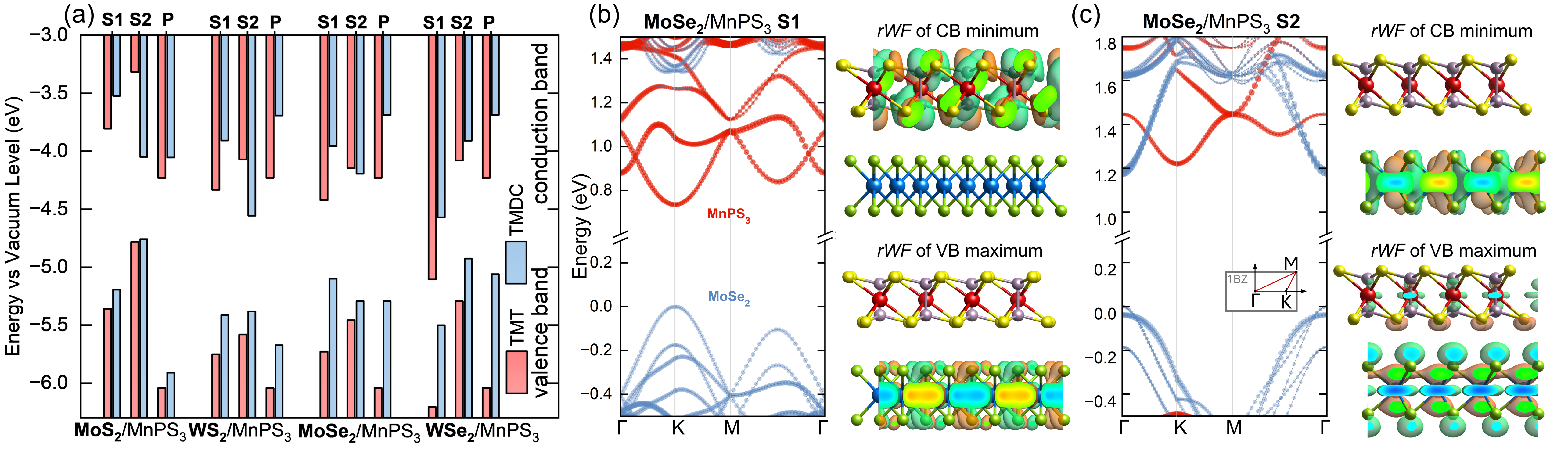}
\caption{(a) Band edge positions (VBM and CBM)  respect to the vacuum level for the employed MnPS$_3$/TMDC heterostructures. Results are shown for two stacking configurations (S1 and S2) as well as for pristine, non-interacting monolayers ("P"), all calculated independently for each system.
(b) Electronic band structure of the MoSe$_2$/MnPS$_3$ heterostructure in the S1 stacking configuration with orbital projections onto the constituent layers (MoSe$_2$ – blue, MnPS$_3$ – red). On the right, the real part of the pseudo-wavefunction (rWF) is visualized for the conduction band minimum (CBM) and valence band maximum (VBM). (c) Band structure and corresponding rWF plots for the same heterostructure in the S2 stacking configuration, showing the spatial character of the CBM and VBM states.}
\label{Fig2}
\end{figure*}

\paragraph{Band alignments.}
Fig.~\ref{Fig2}(a) shows the valence- and conduction-band edge positions referenced to the vacuum level for all systems and stacking configurations, together with the pristine monolayers (P) used as an unstrained reference. Comparing S1 and S2 with P reveals that much of the quantitative variation in band offsets originates from the strain required to construct commensurate supercells. These strain-induced shifts can, in some S2 systems, lead to nominal changes in the apparent alignment type; however, they should not be interpreted as a genuine stacking-controlled modification of the electronic structure. In realistic, nearly incommensurate heterostructures the alignment is expected to remain close to the pristine limit.

In Fig.~S1 and Fig.~S2 of the Supplementary Materials we demonstrate explicitly that the apparent differences between the S1 and S2 band structures originate from the distinct strain patterns imposed by the commensurate supercells and from the Brillouin-zone folding inherent to the 2$\times$2 geometries. These analyses show that the shifted positions of the high-symmetry states, as well as the strain-induced reordering of the MoSe$_{2}$ conduction bands, are purely geometric effects rather than a consequence of interlayer hybridization in the heterostructure.

In the S2 configuration, the $30^\circ$ twist maps the K valley of the primitive TMDC Brillouin zone onto the $\Gamma$ point of the supercell. The VBM and CBM observed at $\Gamma$ in Fig.~\ref{Fig2}(c) are therefore the folded counterparts of the monolayer K-valley states, as confirmed by their layer-resolved wavefunctions. Folding changes only the momentum-space label and does not alter the orbital character or its energetic ordering relative to MnPS$_3$ states.

Across all our calculations, irrespective of material choice, stacking geometry, 
or strain conditions, the valence-band maximum is consistently located in the TMDC 
layer. This result is robust and does not depend on the details of the 
commensurate-cell construction. In contrast, the MnPS$_3$-derived conduction-band states exhibit shifts that we interpret as artefacts of the imposed strain and supercell constraints rather than genuine physical changes. Therefore, while the exact band offsets may vary under these artificial conditions, the qualitative layer character of the band edges—and hence the basic direction of interlayer charge transfer—remains well defined. For this reason, our discussion focuses on the symmetry aspects and spin splittings, which are not affected by these supercell-induced distortions.

\paragraph{Non-relativistic spin splliting.}
\begin{figure*}
\centering
\includegraphics[width=1\textwidth]{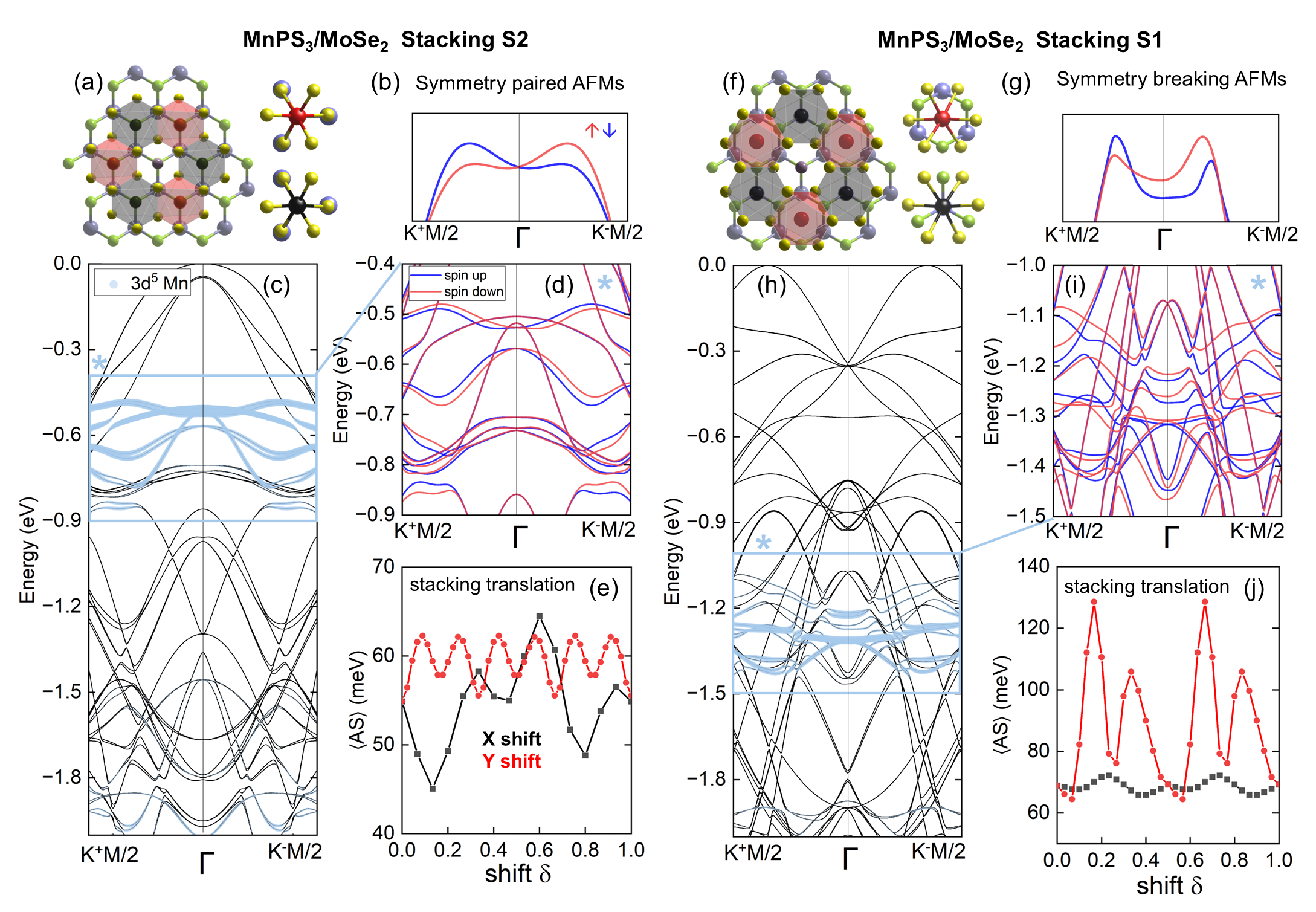}
\caption{Non-relativistic spin splitting (NRSS) and stacking-dependent magnetic effects in MnPS$_3$/TMDC heterostructures. Stacking S2 (shown on the left side of the figure) exhibits symmetry-paired AFM behavior- visualization (a), whereas stacking S1 corresponds to a symmetry-breaking AFM configuration -(f).
(b) Schematic illustration of spin band structures showing two types of antiferromagnetic (AFM) states: symmetry-paired AFMs, characteristic of altermagnetism with opposite spin bands crossing, and (g) symmetry-breaking AFMs. 
 Electronic structure for (c) stacking configuration S2: Spin-unresolved band structure of MnPS$_3$/MoSe$_2$ highlighting the Mn 3d$^5$ contributions; (d) Zoomed-in spin-resolved bands (spin up: blue, spin down: red) along the K$^+$M/2–$\Gamma$–K$^-$M/2 path showing clear altermagnetic splitting;
(e) Average altermagnetic splitting ⟨AS⟩ as a function of in-plane layer translation ($\delta$) for x- and y-shifts.
 Analogous results for stacking configuration S1: (h) Spin-unresolved band structure with Mn 3d$^5$ contributions; (i) Spin-resolved zoom-in revealing NRSS-type splitting without typical spin up/down inversion;
(j) ⟨AS⟩ dependence on stacking translation, showing distinct behavior from the altermagnetic S2 configuration.}
\label{Fig3}
\end{figure*}
The presence of the TMDC layer in the MnPS$_{3}$/TMDC heterostructure leads to significant modifications of the electronic structure compared to the pristine MnPS$_{3}$ monolayer. While bare MnPS$_{3}$ is a collinear Néel-type antiferromagnet, characterized by antiparallel Mn spins on a honeycomb sublattice, its band structure in isolation does not show spin splitting due to the combination of time-reversal and spatial inversion symmetries. However, once interfaced with a TMDC layer, these symmetries are broken by the heterostructure geometry and local environment, allowing nonrelativistic spin splitting (NRSS) to emerge even in the absence of spin-orbit coupling (SOC).
This type of spin splitting has attracted great attention recently and can be broadly classified into two categories, as illustrated schematically in Figure \ref{Fig3}. The first class corresponds to symmetry-paired antiferromagnets, now commonly referred to as altermagnets, where the two spin sublattices are related by a crystal rotation symmetry. In such systems, spin-up and spin-down bands split away from the Brillouin zone center, but must cross at the $\Gamma$ point, where spin degeneracy is protected by rotational symmetry combined with time-reversal (UR symmetry). This leads to alternating spin splitting in momentum space, with bands of opposite spin crossing and to a zero net magnetization. Altermagnetism has been identified in materials such as RuO$_{2}$\cite{doi:10.1126/sciadv.adj4883} and MnTe\cite{Amin2024}, and is a relativistic-symmetry-protected yet SOC-independent phenomenon.
The second class involves symmetry-breaking AFMs, where the two opposite-spin sublattices are not connected by any spatial symmetry. These systems also exhibit NRSS, but unlike altermagnets, they do not maintain spin degeneracy at the $\Gamma$ point. Instead, spin-up and spin-down bands are split throughout the Brillouin zone, including at high-symmetry points. This behavior results from the absence of any operation connecting the spin-structure motif pair, and is captured by the so-called SST-4 symmetry class (Magnetic Space Group type I, without SOC) \cite{Yuan_NRSS}. This type of NRSS is distinct from spin splitting driven by SOC and is purely a result of broken magnetic symmetry.

In our heterostructures, we observe both types of behavior depending on the stacking configuration. Figures \ref{Fig3}(a–j) present detailed band structure analyses for the MnPS$_{3}$/MoSe$_{2}$ system in the two stacking configurations (S1 and S2).
In the S2 configuration, shown in Figures \ref{Fig3}(a–e), the band structure exhibits clear features of altermagnetism. The spin-unresolved band structure in Figure \ref{Fig3}(c) highlights strong contributions from Mn 3$d$ orbitals (light blue), while the spin-resolved zoom-in in panel \ref{Fig3}(d) reveals a characteristic crossing of spin-up and spin-down bands along the K$^+$M/2–$\Gamma$–K$^-$M/2 path. The bands are split away from $\Gamma$, but rejoin at the $\Gamma$ point, indicating symmetry-protected degeneracy - a hallmark of altermagnetism. Importantly, this behavior is sensitive to stacking translation, as shown in Figure \ref{Fig3}(e): the average altermagnetic splitting $\langle \mathrm{AS} \rangle$ varies systematically with relative layer shift, reflecting the influence of local registry on symmetry-breaking perturbations.
By contrast, the S1 configuration, shown in Figures \ref{Fig3}(f–j), exhibits a fundamentally different behavior. In the spin-resolved band structure of Figure \ref{Fig3}(i), the spin-up and spin-down bands remain split even at the $\Gamma$ point, and no clean band inversion is observed. This spin splitting does not correspond to conventional altermagnetism and falls under the NRSS symmetry-breaking class. Figure \ref{Fig3}(j) confirms that the $\langle \mathrm{AS} \rangle$ values in S1 are substantially larger and more asymmetric than in S2, with no signature of symmetry-paired spin alternation. These features indicate that the symmetry linking opposite-spin motifs is fully broken in this configuration.
The distinction arises from the magnetic and structural symmetry of the heterostructures. In the S2 stacking, the 30° rotation between layers preserves a pseudo-rotational symmetry that approximates the conditions for altermagnetism. Although the heterostructure is not strictly invariant under rotation, the spatial pattern of Mn spin motifs remains nearly symmetric, allowing the emergence of band crossings typical of altermagnets. In contrast, the S1 stacking breaks this symmetry completely due to direct alignment of atoms between layers, which imposes a non-equivalent local environment on the spin-up and spin-down Mn sites. This breaks the UR-type protection and leads to non-degenerate spin bands throughout the Brillouin zone - matching the definition of symmetry-breaking NRSS AFMs.
Analyzing the effective magnetic space group (MSG) symmetries provides further insight. For isolated MnPS$_{3}$, the AFM order belongs to a type-III MSG with inversion combined with time reversal. Upon interfacing with a TMDC layer in S2, some of these symmetries survive approximately (e.g., pseudo-C${3}$ rotation), allowing partial symmetry pairing. However, in S1, even these approximate symmetries are lost due to stacking registry, reducing the effective MSG to type-I - the same class responsible for $\Gamma$-split NRSS in ternary nitrides \cite{Yuan_NRSS}.

Taken together, our results show that the stacking geometry directly controls the emergence and nature of NRSS in MnPS$_{3}$/TMDC heterostructures. Depending on the symmetry imposed by stacking, one can selectively induce altermagnetic band crossings or fully split spin bands characteristic of NRSS symmetry-breaking states. This demonstrates a direct link between interlayer symmetry and emergent spin physics in 2D antiferromagnetic heterostructures.

\paragraph{Valley splitting - SOC.}
\begin{figure}[h]
\centering
\includegraphics[width=0.48\textwidth]{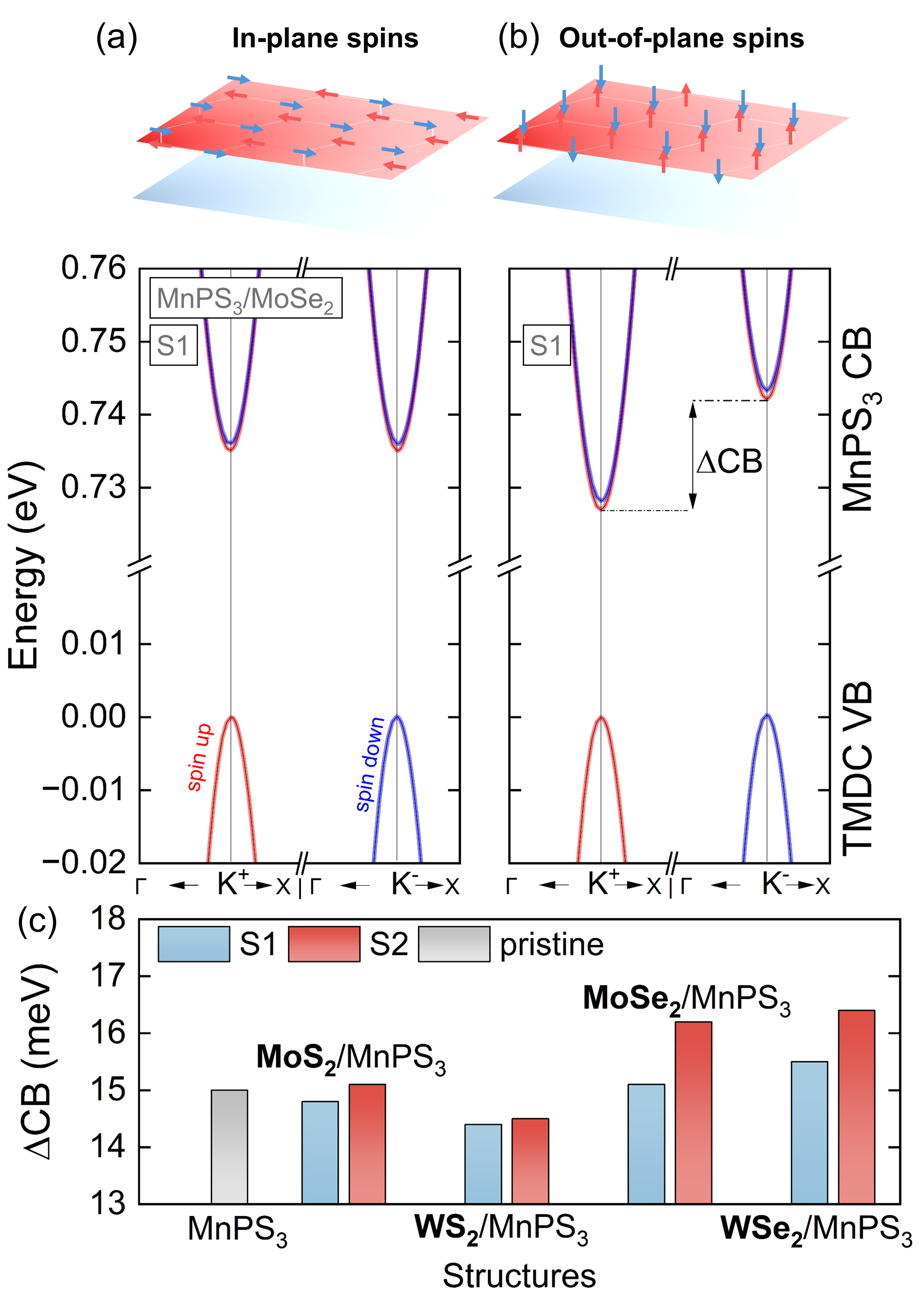}
\caption{Conduction band valley splitting ($\Delta^{CB}$) in various heterostructures as a function of the spin orientation.
Band structures of the MnPS$_3$/MoSe$_2$ in the S1 stacking configuration with Mn spins aligned (a) in-plane and (b) out-of-plane. A conduction band valley splitting ($\Delta^{CB}$) is observed for the out-of-plane spin orientation. The top panels show schematic illustrations of the in-plane and out-of-plane spin orientations.
(c) Comparison of valley splitting $\Delta^{CB}$ for all studied TMDC/MnPS$_3$ heterostructures  and for pristine MnPS$_3$, in both stacking configurations S1 and S2.}
\label{Fig4}
\end{figure}

Once spin–orbit coupling is taken into account, the energies of MnPS$_3$-derived (mainly Mn $3d$) states become sensitive to the orientation of the Mn magnetic moments. Because the conduction-band minimum of the heterostructure is dominated by MnPS$_3$ states, this 
spin-direction dependence directly translates into a valley splitting in the conduction band. Although the TMDC states are only weakly affected, the shifts in the MnPS$_3$ conduction states are sufficient to generate a clear valley-selective response in the optical properties of the combined system.

As illustrated for MnPS$_3$/MoSe$_2$ in Figure~\ref{Fig4}(a), when the Mn spins are aligned in-plane, the valley degeneracy is preserved. In contrast, when the spin axis is oriented out of plane [Figure~\ref{Fig4}(b)], a clear conduction-band splitting $\Delta^{CB}$ appears at the K$^+$ and K$^-$ valleys, originating from the MnPS$_3$ layer, while the TMDC valence band remains nearly unaffected. Notably, the valley splitting occurs in pristine MnPS$_3$. When the Mn spins are rotated from the in-plane to the out-of-plane direction (with SOC included), the degeneracy between the K$^+$ and K$^-$ valleys is lifted, and valley splittings in both the VBM ($\Delta^{VB}$) and CBM ($\Delta^{CB}$) appear, as reported in Ref.~\cite{PhysRevB.109.054426}. This valley splitting arises from the breaking of magnetic symmetries-particularly the combined action of time-reversal symmetry ($\mathcal{T}$) and specific magnetic operations-that otherwise protect valley degeneracy when the spins lie in-plane.

Figure~\ref{Fig4}(c) presents the extracted $\Delta^{CB}$ values for all MnPS$_3$/TMDC heterostructures (MoS$_2$, WS$_2$, MoSe$_2$, and WSe$_2$) and for both S1 and S2 stacking configurations. Although the magnitude of the splitting varies slightly among systems, the effect is consistently present and largely independent of stacking. Our results therefore confirm that the observed valley splitting in MnPS$_3$/TMDC heterostructures originates from the direction of the MnPS$_3$ magnetic moments. Overall, these findings highlight the potential of MnPS$_3$ as a magnetically active 2D substrate capable of controlling valley-selective responses in adjacent TMDC layers. Although the intrinsic magnetic order of MnPS$_3$ remains largely unaffected by stacking or proximity, the spin orientation provides an effective tuning knob for the optical and valleytronic properties of the heterostructure. In particular, rotating the Mn spin orientation can modulate the  magnitude of the valley splitting, enabling magneto-optical control of valley polarization in the TMDC layer.

%\newpage
\section{Conclusions}

In this work, we have performed a comprehensive first-principles study of MnPS$_3$/TMDC heterostructures, examining how stacking geometry, strain, and magnetic order shape their electronic and optical properties. We considered two high-symmetry stacking configurations, S1 and S2, which differ in both interfacial registry and relative twist angle. Due to lattice mismatch, the heterostructures are inherently strained, with the mechanically softer MnPS$_3$ layer accommodating 
a larger part of the deformation.
Although strain and Brillouin-zone folding complicate a direct comparison of the stackings, the key features of the band alignment remain robust. Across all systems, the valence-band maximum resides in the TMDC layer, while the degree of 
interlayer hybridization depends on stacking: weak in S1 and more pronounced in S2. These observations show that the qualitative layer character of the band edges is insensitive to the artificial strain of the commensurate supercell and is therefore reliable.
Spin-resolved band structure analysis uncovers rich symmetry-dependent magnetic effects. In particular, we identify clear signatures of nonrelativistic spin splitting (NRSS) in the absence of spin-orbit coupling, which arise solely from broken spatial symmetries induced by stacking. Depending on the stacking configuration, MnPS$_3$/TMDC heterostructures realize two distinct NRSS regimes.
(i) In S2 stacking configuration, the spin bands display protected crossings away from $\Gamma$, characteristic of altermagnetic order, where spin sublattices are symmetry-related by crystal rotations. (ii) In contrast, the S1 configuration exhibits spin splitting throughout the Brillouin zone, including the $\Gamma$ point, indicative of symmetry-breaking NRSS without paired spin-motifs. This dual behavior demonstrates how stacking geometry can be used to tune the 
magnetic band topology in AFM-based heterostructures.

To quantify magnetic interactions, we extracted exchange and anisotropy parameters from total energy calculations using an extended spin Hamiltonian. The results confirm that interlayer coupling has a limited effect on the magnetism of MnPS$_3$; the observed differences in exchange parameters are mainly due to strain. Anisotropy terms remain small across all systems and stackings, and the magnetocrystalline anisotropy energy ($A$) is largely preserved. This suggests that the fundamental magnetic character of MnPS$_3$ is robust against interfacing with nonmagnetic TMDCs.
Crucially, we demonstrate that magnetism in MnPS$_3$ directly affects the valley physics. When spin-orbit coupling is included and the spin quantization axis is rotated out of plane, we observe significant conduction valley splitting ($\Delta^{CB}$). This splitting results from the breaking of combined time-reversal and mirror symmetries, and is tunable via spin orientation in MnPS$_3$. Our results confirm that valley-selective band responses - previously predicted for pristine MnPS$_3$ - persist in heterostructures and remain robust against stacking and strain variations.
Overall, our study highlights MnPS$_3$ as a versatile 2D antiferromagnetic substrate that can serve both as a symmetry-breaking layer and as a nonvolatile control element for valleytronic and opto-spintronic phenomena in van der Waals heterostructures. By selectively engineering stacking order and magnetic orientation, one can access multiple symmetry regimes - including altermagnetism - and tailor the interplay between spin and valley degrees of freedom. These insights pave the way for future device designs that harness spin-controlled valley selection rules without requiring net magnetization or large spin-orbit coupling.

\section{Acknowledgements}
We thank Thomas Brumme for his support
and fruitful discussions about modelling heterostructures.
Access to computing facilities of the Wroclaw Centre for Networking (WCSS) are gratefully acknowledged. We gratefully acknowledge Polish high-performance computing infrastructure PLGrid (HPC Center: ACK Cyfronet AGH) for providing computer facilities and support within computational grant no. PLG/2025/018673. M.R. thanks National Science Center, Poland for financial support, SONATA 19 Grant 2023/51/D/ST11/02588. M.B. acknowledges financial support from the National Science Centre, Poland under the grant no. 2024/53/B/ST3/04258.

\bibliography{references}
\end{document}

%\tableofcontents
%\section{Electronic parameters}
%\section{Exchange couplings}

\begin{table*}[t]
    \label{tabS1}
    \caption{Structural parameters and band offsets of various MnPS$_3$/TMDC heterobilayers. Optimized lattice constants of heterobilayers ($a_{hBL}$) and pristine monolayers $a_{ML}$ are given in \AA{}. The lattice mismatch ($\varepsilon$) of each layer is quantified by $\epsilon_{ML}=\frac{a_{hBL}-a_{ML}}{a_{ML}} \times 100\%$, where $a_{ML}$ is the optimized lattice parameter of the pristine monolayer. The band offsets are defined as VBO$=$VBM$_{TMDC}-$VBM$_{MnPS_3}$, CBO$=$CBM$_{TMDC}-$CBM$_{MnPS_3}$, in eV.}
    \centering
    \resizebox{\linewidth}{!}{%
    \begin{tabular}{cccccccccc}
        \toprule
        System & Stacking & 
        $a$ (\AA) & $b$ (\AA) & 
        $\varepsilon_\mathrm{TMDC}$ (\%) & 
       $\varepsilon_\mathrm{MnPS_3}$ (\%) & 
        VBM$_{MnPS_3}$(eV) & 
        CBM$_{MnPS_3}$(eV) & 
        VBM$_{TMDC}$(eV) & 
        CBM$_{TMDC}$(eV) \\
        \midrule
        \multirow{2}{*}{MnPS$_3$/MoS$_2$} & S1 & 6.245 & 10.817 & 1.4 &  3.0 & -5.36 & -3.81 & -5.20 & -3.52 \\
                                 & S2 & 5.714 &  9.895 &  4.2 & -5.8 & -4.78 & -3.32 & -4.76 & -4.05 \\
        \multirow{2}{*}{MnPS$_3$/WS$_2$}  & S1 & 6.262 & 10.847 & -1.4 &  3.3 & -5.75 & -4.33 & -5.41 & -3.91 \\
                                 & S2 & 5.715 &  9.898 &  3.9 & -5.8 & -5.58 & -4.07 & -5.38 & -4.56 \\
        \multirow{2}{*}{MnPS$_3$/MoSe$_2$}& S1 & 6.417 & 11.114 & -2.7 &  5.8 & -5.73 & -4.42 & -5.10 & -3.96 \\
                                 & S2 & 5.857 & 10.402 &  1.0 & -3.4 & -5.46 & -4.15 & -5.30 & -4.20 \\
        \multirow{2}{*}{MnPS$_3$/WSe$_2$} & S1 & 6.421 & 11.122 & -2.5 &  5.9 & -6.21 & -5.11 & -5.20 & -4.57 \\
                                 & S2 & 5.847 & 10.127 &  2.5 & -3.6 & -5.30 & -4.08 & -4.93 & -3.91 \\
        \hline
        MnPS$_3$ &  & 6.064 & 10.504 & -- & -- & -6.05 & -4.23 & -- & -- \\
        MoS$_2$  &  & 3.167 & 5.486     & -- & -- & --   & --   & -5.91 & -4.05 \\
        WS$_2$   &  & 3.176 & 5.501     & -- & -- & --   & --   & -5.68 & -3.69 \\
        MoSe$_2$ &  & 3.299 & 5.713     & -- & -- & --   & --   & -5.30 & -3.69 \\
        WSe$_2$  &  & 3.294 & 5.706     & -- & -- & --   & --   & -5.06 & -3.34 \\
        \bottomrule
    \end{tabular}
    }
\end{table*}

\begin{table*}[t]
\caption{Heisenberg exchange parameters ($J_1$, $J_2$, $J_3$) and anisotropy constants ($\lambda_1$, $\lambda_2$, $\lambda_3$, $A$) for MnPS$_3$/TMDC heterostructures in various S1 stacking configurations. All values are given in meV. The negative values $J_i$ indicate the AFM exchange couplings, the positive value of $A$ denotes the preferred out-of plane magnetization axis. }
\centering
\resizebox{\linewidth}{!}{%
\begin{tabular}{c c c c c c c c c c}
TMT & TMDC & stacking & $J_1$ & $J_2$ & $J_3$ & $\lambda_1$ & $\lambda_2$ & $\lambda_3$ & $A$ \\
\hline
\multirow{14}{*}{MnPS$_3$} & --    & pure ($a{=}6.06$ \AA{})   & -0.74 & -0.06 & -0.31 & $-2.69 \times 10^{-3}$ & $1.14 \times 10^{-3}$ & $-8.60 \times 10^{-4}$ & $3.96 \times 10^{-3}$ \\
                           & --    & stressed as MoS$_2$ & -0.57 & -0.06 & -0.28 & $-1.62 \times 10^{-3}$ & $9.81 \times 10^{-4}$ & $-7.15 \times 10^{-4}$ & $4.08 \times 10^{-3}$ \\
\cline{3-10}
 & \multirow{3}{*}{MoS$_2$} & S1$_0$ & -0.53 & -0.04 & -0.26 & $7.11 \times 10^{-4}$  & $3.94 \times 10^{-4}$ & $3.53 \times 10^{-4}$  & $2.94 \times 10^{-3}$ \\
 &                          & S1$_1$ & -0.54 & -0.04 & -0.26 & $1.07 \times 10^{-3}$  & $4.80 \times 10^{-4}$ & $3.78 \times 10^{-4}$  & $3.29 \times 10^{-3}$ \\
 &                          & S1$_2$ & -0.54 & -0.04 & -0.26 & $3.20 \times 10^{-4}$  & $5.70 \times 10^{-5}$ & $6.34 \times 10^{-5}$  & $3.05 \times 10^{-3}$ \\
\cline{3-10}
 & \multirow{3}{*}{WS$_2$}  & S1$_0$ & -0.52 & -0.04 & -0.26 & $-1.11 \times 10^{-3}$ & $-7.31 \times 10^{-4}$ & $-5.35 \times 10^{-4}$ & $3.74 \times 10^{-3}$ \\
 &                          & S1$_1$ & -0.52 & -0.04 & -0.26 & $-1.30 \times 10^{-3}$ & $-8.68 \times 10^{-5}$ & $-5.92 \times 10^{-5}$ & $8.03 \times 10^{-3}$ \\
 &                          & S1$_2$ & -0.52 & -0.04 & -0.26 & $-9.49 \times 10^{-4}$ & $-3.48 \times 10^{-5}$ & $-3.52 \times 10^{-5}$ & $4.12 \times 10^{-3}$ \\
\cline{3-10}
 & \multirow{3}{*}{MoSe$_2$}& S1$_0$ & -0.37 & -0.03 & -0.21 & $-6.90 \times 10^{-5}$ & $-3.56 \times 10^{-3}$ & $-2.37 \times 10^{-3}$ & $6.74 \times 10^{-3}$ \\
 &                          & S1$_1$ & -0.37 & -0.03 & -0.22 & $3.72 \times 10^{-4}$  & $9.16 \times 10^{-5}$  & $2.91 \times 10^{-5}$  & $3.10 \times 10^{-3}$ \\
 &                          & S1$_2$ & -0.37 & -0.03 & -0.22 & $-2.98 \times 10^{-3}$ & $-1.59 \times 10^{-3}$ & $-1.05 \times 10^{-3}$ & $3.55 \times 10^{-3}$ \\
\cline{3-10}
 & \multirow{3}{*}{WSe$_2$} & S1$_0$ & -0.37 & -0.03 & -0.21 & $3.28 \times 10^{-4}$  & $1.16 \times 10^{-4}$  & $-7.67 \times 10^{-4}$ & $4.36 \times 10^{-3}$ \\
 &                          & S1$_1$ & -0.36 & -0.03 & -0.22 & $-1.19 \times 10^{-4}$ & $-1.91 \times 10^{-4}$ & $-1.01 \times 10^{-4}$ & $3.32 \times 10^{-3}$ \\
 &                          & S1$_2$ & -0.37 & -0.03 & -0.21 & $-8.58 \times 10^{-5}$ & $-1.16 \times 10^{-4}$ & $-1.23 \times 10^{-4}$ & $3.54 \times 10^{-3}$ \\
\end{tabular}
}
\label{tabS2}
\end{table*}

%\bibliography{references}

\begin{figure*}\centering
\includegraphics[width=1\textwidth]{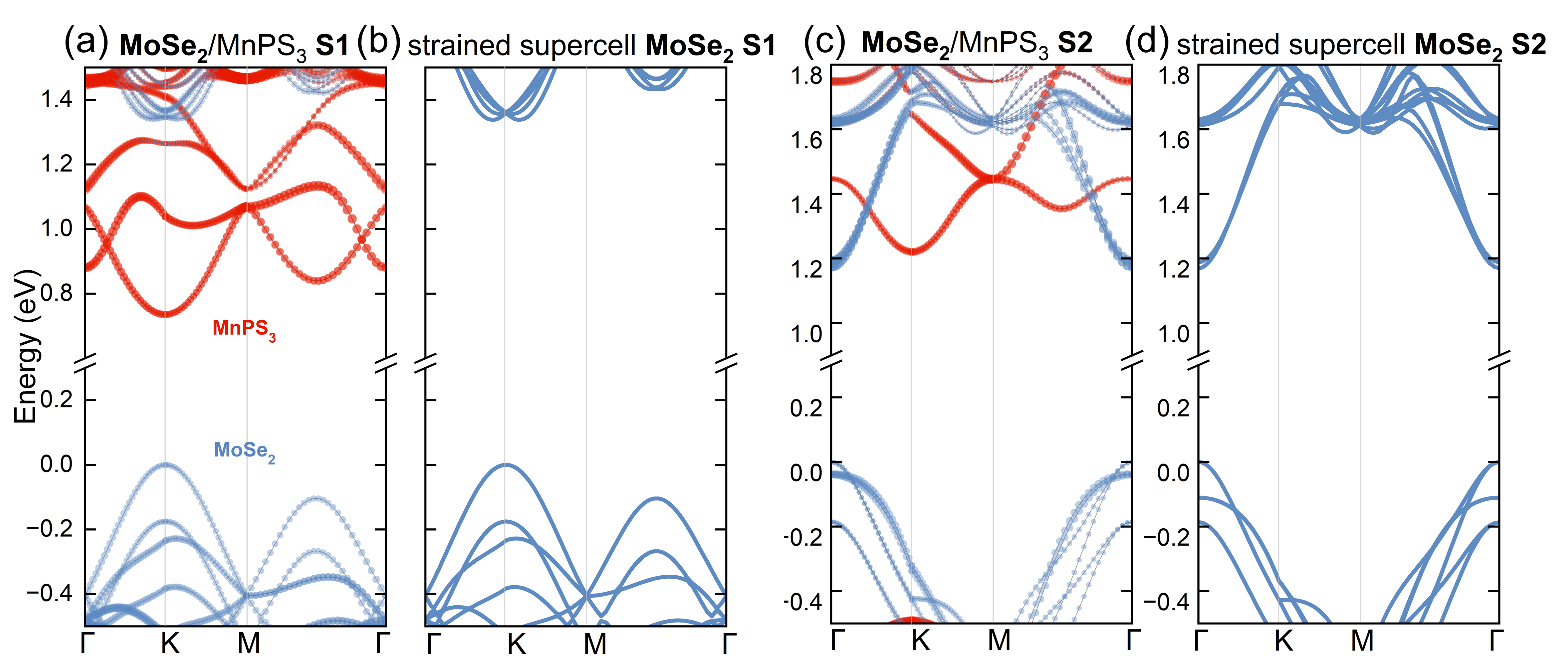}
\caption{Electronic band structures of MnPS$_3$/MoSe$_2$ heterostructures and strained MoSe$_2$ monolayers for stackings S1 and S2.
(a,c) Band structures of the MnPS$_3$/MoSe$_2$ heterostructure for stackings S1 and S2, with orbital projections onto MoSe$_2$ (blue) and MnPS$_3$ (red). (b,d) Band structures of isolated MoSe$_2$ computed in the same strained supercells as in (a,c), obtained by removing the MnPS$_3$ layer while keeping the lattice parameters fixed.
Comparison between the heterostructure and the strained MoSe$_2$ clearly shows that the dominant source of the modifications in the MoSe$_2$ bands is strain imposed by the commensurate supercell, not the proximity to MnPS$_3$. For S1, the MoSe$_2$ band structure is nearly identical to its strained counterpart, confirming very weak interlayer hybridization. For S2, noticeable differences appear particularly in the valence band region, reflecting enhanced interlayer interaction and consistent with the real-space wavefunction analysis in Figure 3 of the main text.
Overall, the figure demonstrates that strain accounts for most of the band-structure changes, while proximity effects are strongly stacking-dependent and become significant primarily in the S2 configuration.}
\label{Fig1_s}
\end{figure*}

\begin{figure*}\centering
\includegraphics[width=1\textwidth]{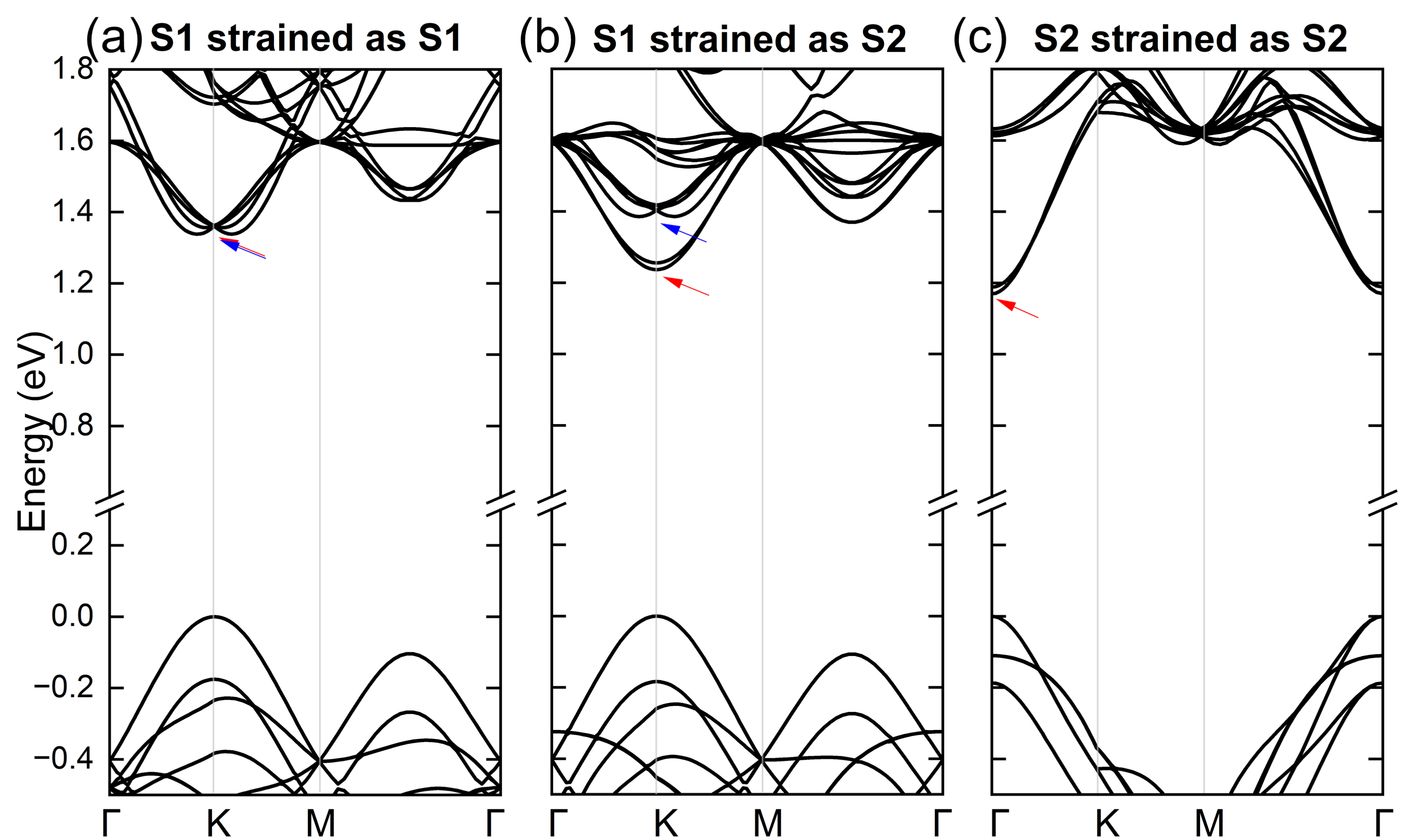}
\caption{ Electronic band structures of monolayer MoSe$_{2}$ calculated under different strain and supercell conditions, illustrating the role of strain and Brillouin-zone folding for stacking configurations S1 and S2. 
(a) MoSe$_{2}$ in the S1 supercell under the strain corresponding to the S1 heterostructure ("S1 strained as S1"). 
(b) MoSe$_{2}$ placed in the S1 supercell but kept under the strain of S2 ("S1 strained as S2"). 
(c) MoSe$_{2}$ fully strained and folded as in S2 ("S2 strained as S2"). 
The comparison shows that the differences between the S1 and S2 band structures arise from two effects: 
(i) distinct Brillouin-zone folding in the 2$\times$2 supercells used for stackings S1 and S2, which shifts the high-symmetry states to different k-points, and 
(ii) strain-induced band-pressure effects, which in the S1 geometry cause a reordering of the conduction-band states (blue and red arrows). 
The test calculation "S1 strained as S2" confirms that the apparent S1/S2 differences originate from the strain pattern and supercell folding rather than from interlayer interactions. 
In the MnPS$_{3}$/MoSe$_{2}$ heterostructures the conduction-band minimum is provided by MnPS$_{3}$, so this strain-driven reordering of the MoSe$_{2}$ conduction bands does not affect the interpretation of the band alignment. Under realistic low-strain experimental conditions, MoSe$_{2}$ would not exhibit such conduction-band reordering, confirming that the observed S1/S2 differences reflect only the structural constraints of the commensurate supercell.
}
\label{Fig2_s}
\end{figure*}